\documentclass[runningheads]{llncs}
\usepackage{amsmath}
\usepackage{graphicx}
\usepackage{hyperref}
\usepackage[final]{listings}
\usepackage{xargs}
\usepackage[dvipsnames]{xcolor}
\usepackage[colorinlistoftodos,prependcaption,textsize=tiny]{todonotes}
\newcommandx{\georg}[2][1=]{\todo[prepend,caption={G},linecolor=red,backgroundcolor=red!25,bordercolor=red,#1]{#2}}
\newcommandx{\viktor}[2][1=]{\todo[prepend,caption={V},linecolor=blue,backgroundcolor=blue!25,bordercolor=blue,#1]{#2}}
\newcommandx{\antoine}[2][1=]{\todo[prepend,caption={A},linecolor=green,backgroundcolor=green!25,bordercolor=green,#1]{#2}}
\lstdefinelanguage{console}{
  morekeywords={,invalid,},
  basicstyle={\footnotesize\sffamily\ourlstcolor},
  keywordstyle={\color{red}\bfseries},
}

\lstdefinelanguage{scala}{
  alsoletter={@},
  morekeywords={, abstract, case, catch, choose, class, def, do, else, extends, final, finally, for, if, implicit, import, match, new, null, object, let, in, be, lazy,holds,@erasable,
override, package, private, protected, requires, ensures, decreases, return, sealed, super, then, this, throw, trait, try, type, val, var, while, yield, domain, template, assume, fun,
postcondition, precondition,invariant, constraint, assert, each, _, return, @generator, @ghost, ghost, @opaque, @extern, ensure, require, ensuring, assuming, asserting, reads,
modifies, check, }
  sensitive=true,
  morecomment=[l]{//},
  morecomment=[s]{/*}{*/},
  morestring=[b]"
}
\definecolor{verydarkblue}{RGB}{0,20,80}
\definecolor{verydarkgreen}{RGB}{0,80,20}
\newcommand{\ourlstcolor}{\color{verydarkblue}}
\lstMakeShortInline[columns=fullflexible]| 

\newcommand{\RA}{\Rightarrow}

\newcommand{\rA}{\rightarrow}

\begin{document}
\sloppy
\title{Proving and Disproving Programs with \\ Shared Mutable Data}
%
%
%
%
\author{Georg Schmid \and Viktor Kunčak}
\institute{LARA, EPFL, Switzerland\\
\email{\{georg.schmid, viktor.kuncak\}@epfl.ch}}
\maketitle              
\begin{abstract}
We present a tool for verification of deterministic programs with shared mutable references against specifications such as assertions, preconditions, postconditions, and read/write
effects.  We implement our tool by encoding programs with mutable references into annotated purely functional recursive programs. We then rely on function unfolding and the SMT solver Z3 to prove or disprove safety and to establish program termination.  Our tool uses a new translation of programs where frame conditions are encoded using quantifier-free formulas in first-order logic (instead of relying on quantifiers or separation logic).  This quantifier-free encoding enables SMT solvers to prove safety or report counterexamples relative to the semantics of procedure specifications.  Our encoding is possible thanks to the expressive power of the extended array theory of the Z3 SMT solver.  In addition to the ability to report counterexamples, our tool retains efficiency of reasoning about purely functional layers of data structures, providing expressiveness for mutable data but also a significant level of automation for purely functional aspects of software.  We illustrate our tool through examples manipulating mutable linked structures and arrays.
%
\footnote{This manuscript is submitted for peer review to CAV 2021 in January 2021 and may only be used to facilitate peer review process
for this conference.}

\end{abstract}

\section{Introduction}

Formal verification of programs with shared mutable data
structures is a long-standing problem.  Among the most
promising techniques used in today's verification tools are
separation logic \cite{DBLP:conf/csl/OHearnRY01} with bi-abduction
\cite{DBLP:journals/jacm/CalcagnoDOY11}
implemented in the Infer tool \cite{DBLP:journals/cacm/DistefanoFLO19} used by Facebook
and dynamic frames \cite{DBLP:conf/fm/Kassios06} realized in
tools including Dafny \cite{DBLP:conf/lpar/Leino10b} and used
to verify complex software systems at Microsoft \cite{DBLP:conf/sosp/HawblitzelHKLPR15}. Separation logic and dynamic
frames are closely related, as demonstrated by the
approach in the VeriFast tool
\cite{DBLP:journals/toplas/SmansJP12}.

This paper presents a tool
for
reasoning about mutable programs in a subset of the Scala
programming language \cite{Odersky19ScalaBook}.
Like the dynamic frames approach, we use defined or constrained sets of objects to specify
frame conditions.
Like Dafny, our tool uses SMT solvers to
establish properties instead of dedicated symbolic execution
for heap-manipulating programs in several other
approaches. We also model the heap as a function from storage
locations to values.

However, our encoding of frame conditions is different
from the one in Dafny.
Unlike Dafny, which reduces programs
to a guarded-command language Boogie \cite{DBLP:conf/fmco/BarnettCDJL05},
we reduce our programs to recursive functional programs that manipulate
rich data types supported by
the Z3 SMT solver \cite{DBLP:conf/tacas/MouraB08}, building on top of the publicly available Stainless
verification infrastructure for functional programs \cite{HamzaETAL19SystemFR}.
Whereas the encoding in Dafny makes use of \emph{universal quantifiers with triggers} to encode frame conditions
(expressing that \emph{all} non-modified locations remain the same),
we avoid quantifiers and instead use a \emph{generalized theory of arrays} \cite{DBLP:conf/fmcad/MouraB09} of
Z3. This expressive theory of arrays comes with completeness guarantees for
satisfiability checking of quantifier-free formulas combining arrays and other
theories using model-based theory combination \cite{DBLP:journals/entcs/MouraB08}.
Thanks to this new encoding and the decision procedures of Z3, our verification tool can report meaningful counterexamples for invalid
properties, even in the cases where the bodies of methods are abstracted by their modifies clauses.
In contrast, SMT solvers typically cannot report counterexamples to satisfiability for formulas with universal quantifiers.

The fact that our tool reduces verification conditions to functional programs also allows users of our tool
to leverage the expressive power of recursive functional programming in implementation and specification.
Our tool need not encode immutable algebraic data types using the heap nor add heap arguments to pure functions.
The result is a better
verification experience on a mix of purely functional and mutable code, compared to a more uniform encoding.

In this paper we use a running example to outline how to use our tool to specify and verify Scala programs that contain mutable data.
We then sketch our encoding into recursive functions via the extended array theory \cite{DBLP:conf/fmcad/MouraB09} and
discuss our experience with the tool on verifying shared mutable data structures.

\begin{figure}[htbp]
  \label{code:treemapnospec}
\begin{lstlisting}[numbers=left,basicstyle={\footnotesize\sffamily\ourlstcolor}]
case class Cell[T](var value: T) extends AnyHeapRef

case class Leaf[T](data: Cell[T]) extends Tree[T]
case class Branch[T](left: Tree[T], right: Tree[T]) extends Tree[T]
sealed abstract class Tree[T] {
  @ghost def repr: Set[AnyHeapRef] = this match { // all cells in the tree
    case Leaf(data) $\RA$ Set[AnyHeapRef](data)
    case Branch(left, right) $\RA$ left.repr ++ right.repr
  }

  def tmap(f: T $\RA$ T): Unit = { // minimal specification
    reads(repr)
    modifies(repr)
    decreases(this)

    this match {
      case Leaf(data)          $\RA$ data.value = f(data.value)
      case Branch(left, right) $\RA$ left.tmap(f); right.tmap(f)
    }
  }
}
\end{lstlisting}
  \caption{A tree with mutable leaves and a parallelizable in-place map, including read and write frame conditions and a termination measure.}
\end{figure}

\section{Example}
\label{sec:example}

To illustrate a mix of functional and imperative programming, Figure~\ref{code:treemapnospec}
shows a binary tree data structure whose interior nodes are immutable but whose leaves store
mutable references of generic type |T|.
The code is accepted by the standard Scala 2.12 compilation pipeline given
appropriate library imports.
Our tool supports a fragment of Scala with functional features (such as pure first-class functions)
as well as imperative features (mutable fields) and object-oriented features (traits and dynamic dispatch).
For any class, users explicitly opt into mutability and heap reasoning by inheriting from |AnyHeapRef|.
Our focus is the method  |def tmap(f: T $\RA$ T)| on the |Tree| class, which applies an in-place transformation |f| to all leaf cells.
For example, given a |tree: Tree[BigInt]|, invoking |tree.tmap(n $\RA$ n + 1)| increments the values in all the
leaves of |tree| by one.
The method recursively traverses the tree and updates all cells upon reaching the leaves. As written, the code is sequential.
That said, when all leaf cells are distinct, the updates in subtrees can all proceed in parallel, without race conditions, resulting in
logarithmic algorithm depth \cite{10.1145/227234.227246} for balanced trees.

Figure~\ref{code:treemapnospec} is also a minimally specified program accepted by our tool, which automatically
verifies the conformance of |tmap| to its declared effects.
The |reads| clause indicates that the only mutable references that |tmap| reads are given
by the value returned from auxiliary function |repr|, which computes the set of mutable cells
in a given tree. Similarly, |modifies| indicates that these are the only sets the method
is allowed to modify, which means that all other mutable objects remain the same after a call to |tmap|.
Finally, the |decreases| clause provides an explicit decreasing measure on the size of the receiver
object of |tmap| that enables the system to show that
the function terminates. (For the simpler function |repr|, the system automatically infers the termination measure.)
The |@ghost| annotation ensures that the |repr| function is not accidentally executed, but can only be used in specifications
that are erased at run time.

If we try to omit a reads or modifies clause, or incorrectly define |repr| to not descend into subtrees, the tool reports
a counterexample state detecting that the specification |reads| or |modifies| is violated, with a message such as
\begin{lstlisting}[language=console]
tmap  body assertion: reads of TreeImmutMapGenericExample.Tree.tmap   invalid

\end{lstlisting}
pointing to an undeclared effect in line 18 of Figure~\ref{code:treemapnospec}.

Next, consider the following test method:
\begin{lstlisting}[numbers=none,basicstyle={\footnotesize\sffamily\ourlstcolor}]
def test(t: Tree[Int], c: Cell[Int]) = {
  require(c.value == 0)
  reads(t.repr ++ Set[AnyHeapRef](c))
  modifies(t.repr)

  t.tmap(x $\RA$ (x | 1))
} ensuring(_ $\RA$ c.value == 0)
\end{lstlisting}

If we mark |tmap| using the |@opaque| annotation to prevent function unfolding and
try to verify |test|, the system reports a counterexample, such as this one:

\begin{lstlisting}[language=console]
Found counter-example:
   t: Tree[Int]                $\rA$ Leaf[Int](HeapRef(4))
   c: HeapRef                  $\rA$ HeapRef(4)
   heap0: Map[HeapRef, Object] $\rA$ {HeapRef(4) $\rA$ Cell(Cell[Object](SignedBitvector32(0))), * $\rA$ SignedBitvector32(2)}
\end{lstlisting}
indicating that, when |tmap| is approximated with its effects,
the |ensuring| clause can be violated when tree |t| contains precisely the reference |c|.

\begin{figure}[hbtp]
  \label{code:treemap}
\begin{lstlisting}[numbers=left,basicstyle={\footnotesize\sffamily\ourlstcolor}]
def tmap(f: T $\RA$ T): Unit = { // strong specification
  reads(repr)
  modifies(repr)
  require(valid)
  @ghost val oldList = toList

  this match {
    case Leaf(data) $\RA$
      data.value = f(data.value)
      ghost { check(toList == oldList.map(f)) }

    case Branch(left, right) $\RA$
      @ghost val (oldList1, oldList2) = (left.toList, right.toList)
      assert(left.repr $\cap$ right.repr == $\emptyset$)
      left.tmap(f); right.tmap(f)
      ghost {
        lemmaMapConcat(oldList1, oldList2, f)
        check(toList == oldList.map(f))
      }
  }
} ensuring (_ $\RA$ toList == old(toList.map(f)))  // main property

def valid: Boolean =  // tree invariant: subtrees store disjoint cells
  this match {
    case Leaf(data) $\RA$ true
    case Branch(left, right) $\RA$
      left.repr $\cap$ right.repr == $\emptyset$ &&
      left.valid && right.valid
  }

def toList: List[T] = { // abstraction function
  reads(repr)
  this match {
    case Leaf(data) $\RA$ List(data.value)
    case Branch(left, right) $\RA$ left.toList ++ right.toList
  }
}

def lemmaMapConcat[T, R](xs: List[T], ys: List[T], f: T $\RA$ R): Unit = {
  xs match {
    case Nil() $\RA$ ()
    case Cons(_, xs) $\RA$ lemmaMapConcat(xs, ys, f)
  }
} ensuring (_ $\RA$ xs.map(f) ++ ys.map(f) == (xs ++ ys).map(f))
\end{lstlisting}
  \caption{Functional correctness of the {\protect \lstinline{tmap}} method including the abstraction function, the invariant, and a proven
    lemma about purely functional lists. We use $\cap$ to display intersection of sets {\tt \&},
    and use $\emptyset$ for the empty set of heap references {\tt Set[AnyHeapRef]()}. The {\tt ++} symbol denotes
  concatenation of functional lists, as in Scala.}
\end{figure}

To illustrate specification of stronger correctness properties, we show that |tmap| behaves
like |map| on purely functional lists.
This stronger specification of |tmap| is in the |ensuring| (postcondition) clause of the version of |tmap| in Figure~\ref{code:treemap}
(line 21). The property is interesting because it gives us assurance of
correctness while being able
to write code that reuses memory locations and permits parallelization.
The property is expressed by defining an abstraction function \cite{DBLP:journals/tcs/AbadiL91} |toList|
that maps the tree into the sequence of elements stored in its leaf cells.
(The purely functional |List| data type and the |map| function on lists are defined in the standard library of Stainless.)
To prove the |ensuring| clause, it is necessary to introduce a precondition for |tmap|, expressed
using the construct |require(valid)|.
The |valid| method returns true when all subtrees store disjoint cells.
The |tmap| method may then only be called when this predicate holds.
The assertion on line 14 follows directly from |valid| and
expresses disjointness of the side effects of calls on line 15, which
makes it easy for a program transformation to safely parallelize these calls \cite{DBLP:journals/tpds/HendrenN90}.
Remarkably, although it expresses separation,
the |valid| method in this implementation does not depend on the content of mutable cells, but only on the identity of references.
Our tool checks this independence thanks to the absence of |reads| and |modifies| clauses in the signature of |valid|.
Because it does not depend on mutable state, |valid| trivially continues to hold after |tmap| is invoked.

In many cases our tool can automatically prove properties of interest thanks to SMT solvers and the unfolding algorithm of Stainless.
On the other hand, showing complex properties such as functional correctness in this case may require different types of hints.
If the size and the complexity of SMT formulas overwhelm the solver the user can use auxiliary assertions,
expressed using |assert| and |check| functions in Figure~\ref{code:treemap}. The system also requires
guidance for inductive properties when reasoning does not follow the pattern of functions that are iteratively unfolded. In such
cases, we need to introduce lemmas and prove them using recursion to express inductive arguments, as with
|lemmaMapConcat| defined in lines 39-44 and instantiated in line 17.
With these specifications and hints in place, our tool successfully verifies the functional correctness and termination of |tmap| in
a matter of seconds.

\section{Encoding}
\label{sec:encoding}

\begin{figure}[btp]
\begin{lstlisting}[numbers=left,basicstyle={\footnotesize\sffamily\ourlstcolor}]
case class HeapRef(id: BigInt)

case class Cell$_\text{Data}$[T](value: T)

sealed abstract class Tree[T]
case class Leaf[T](data$_\text{ref}$: HeapRef) extends Tree[T]
case class Branch[T](left: Tree[T], right: Tree[T]) extends Tree[T]
\end{lstlisting}
  \caption{The data types of the running example in Section~\ref{sec:example} after our encoding.}
  \label{code:treemapencodedtypes}
\end{figure}

We briefly describe our heap encoding and how it achieves framing without quantification.
Our tool builds upon the existing counterexample-complete unfolding procedure of the Stainless verifier and
exploits the additional expressive power afforded by combinatory array logic \cite{DBLP:conf/fmcad/MouraB09}, an extended array theory available in Z3.
This use of array combinators for framing is, to the best of our knowledge, novel.
Notably, our encoding allows for a high degree of proof automation without giving up counterexamples.

\paragraph{Heap Representation.}
Our tool models stateful operations by explicitly reading from and updating a global, mutable map that relates each object to its state.
In a second transformation step such programs with local mutations are reduced to functional ones:
We add an explicit, immutable heap parameter to each procedure and return a new, potentially updated heap along with the regular outputs.
Stateful operations such as reads and writes within the procedure are thus replaced by pure operations on a map.
Given our context of Scala, a statically-typed language, the heap is internally modelled by a map |heap| of type |Heap = Map[HeapRef, Any]|
where |Any| is the top type and |HeapRef| is a data type representing an object's identity.

\begin{figure}[hbt]
\begin{lstlisting}[numbers=left,basicstyle={\footnotesize\sffamily\ourlstcolor}]
def tmap[T](h0: Heap, t: Tree[T], f: T $\RA$ T): (Unit, Heap) = {
  val (rs, ms) = (repr(t), repr(t))
  t match {
    case Leaf(data$_\text{ref}$) $\RA$
      assert(data$_\text{ref}$ $\in$ rs, "`data` must be in reads set")
      assert(data$_\text{ref}$ $\in$ ms, "`data` must be in modifies set")
      val data: Cell$_\text{Data}$ = {
        assume(h0(data$_\text{ref}$).isInstanceOf[Cell$_\text{Data}$[T]])
        h0(data$_\text{ref}$).asInstanceOf[Cell$_\text{Data}$[T]]
      }
      val data': Cell$_\text{Data}$ = Cell$_\text{Data}$(f(data.value))
      ((), h0.updated(data$_\text{ref}$, data'))

    case Branch(left, right) $\RA$
      val (_, h1) = tmap$_\text{shim}$(h0, rs, ms, left, f)
      tmap$_\text{shim}$(h1, rs, ms, right, f)
  }
}

def tmap$_\text{shim}$[T](h0: Heap, rd: RSet, md: RSet, t: Tree[T], f: T $\RA$ T): (Unit, Heap) = {
  val (rs, ms) = (repr(t), repr(t))
  assert(rs $\subseteq$ rd, "reads set of Tree.tmap")
  assert(ms $\subseteq$ md, "modifies set of Tree.tmap")
  val res @ (_, h1) = tmap(h0, t, f)
  assume(res == tmap(rs.mapMerge(h0, dummyHeap), t, f))
  assume(h1 == ms.mapMerge(h1, h0))
  res
}
\end{lstlisting}
  \caption{The result of encoding the minimally-specified {\protect \lstinline{tmap}} method of Section~\ref{sec:example}.
  We use $\subseteq$ to typeset {\protect \lstinline{subsetOf}}, $\in$ for {\protect \lstinline{contains}}, and
  abbreviate {\protect \lstinline{Set[AnyHeapRef]}} by {\protect \lstinline{RSet}}.}
  \label{code:treemapencodedfun}
\end{figure}

\paragraph{Encoding {\protect \lstinline{tmap}}.}
We will explain our encoding by the example of the minimally-specified version of |tmap| on |Tree| (cf.\ Figure~\ref{code:treemapnospec}).
In Figure~\ref{code:treemapencodedtypes} we show the data types after transformation.
We treat \emph{heap types}, i.e., descendants of |AnyHeapRef|, like |Cell|, differently from immutable types such as |Tree|.
The latter are translated into algebraic data types in the obvious way (lines 5-7).
References to heap types, on the other hand, are erased to the internal ADT |HeapRef| that is isomorphic to the natural numbers (line 1).
For instance, the field |data: Cell[T]| of |Leaf| becomes |data$_\text{ref}$: HeapRef| (line 6).
Additionally, each heap class like |Cell| is translated to a single-constructor ADT that encapsulates an object's state at a given time, e.g., |Cell$_\text{Data}$| (line 3).

In Figure~\ref{code:treemapencodedfun} we show the encoding of |tmap| itself.
The method is reduced to a type-parametric function that takes its original argument |f|, the method receiver |t| and a heap parameter |h0|.
The imperative operations in |tmap| are translated to functional operations on |Heap| as mentioned above, and the modified heap is returned along with the original return value.
In particular, if the current tree |t| is a leaf, then we extract its reference to a cell |data$_\text{ref}$| (line 4) and
index the initial heap |h0| at |data$_\text{ref}$| (line 9).
Note that since the heap map stores values of type |Any| we have to perform a downcast (lines 8-9).
This is safe, since we will only verify well-typed Scala programs, so any such cast will be correct by construction.
On line 11 we apply the function |f| to the old |value| of |data| and construct a |Cell$_\text{Data}$| value reflecting the new state of |data|.
We then return the updated heap on line 12.
In case the tree |t| is a |Branch| we simply perform two recursive calls (lines 15-16), albeit through the newly-introduced wrapper function |tmap$_\text{shim}$|.

Our encoding achieves modular verification of heap contracts (|reads| and |modifies|) by injecting some additional assertions and assumptions.
We bind the |reads| and |modifies| sets (|rs| and |ms|) at the top of the function (line 2).
For each object that is read or modified we check that the object is in the respective set (lines 5-6).
For function calls we check that the callee's |reads|, resp.\ |modifies|, set is subsumed by the caller's.
We achieve this by invoking a wrapper function |tmap$_\text{shim}$|, that additionally takes as parameters the \emph{domains} on which the passed heap is defined for
reads and modifications (|rd| and |md|).
Within the wrapper we bind the original function's |reads| and |modifies| sets (line 21),
check subsumption wrt.\ the domains (lines 22-23) and call the original function |tmap| (line 24).
Finally, we also add the modular guarantees about |tmap| as assumptions:
namely, that the result of |tmap| only depends on the |reads| subset of the heap (line 25), and
that the heap resulting from |tmap| may only have changed on objects in |modifies| (line 26).
These two assumptions are encoded using the \emph{{\protect \lstinline{mapMerge}} primitive},
which can be seen as a ternary operator of type |$\forall$ K V. Set[K] $\RA$ Map[K,V] $\RA$ Map[K,V] $\RA$ Map[K,V]|.
Specifically, |mapMerge| takes a set |s| along with two maps |m1|, |m2| and produces a map |m' = s.mapMerge(m1, m2)| such that
|$\forall$ k:K. (k $\in$ s $\rA$ m'[k] = m1[k]) $\wedge$ (k $\not\in$ s $\rA$ m'[k] = m2[k])|.

\paragraph{Quantifier-Free Frame Conditions.}
For our encoding of frame conditions we leverage the fact that Stainless translates both sets and maps to the theory of (infinite, extensional) arrays in Z3.
This means that reads and modifies expressions of type |Set[AnyHeapRef]| become arrays typed |HeapRef $\RA$ Boolean|,
while heap maps of type |Map[HeapRef, Any]| are translated to |HeapRef $\RA$ Any|.
We can then use the array combinator $\text{map}_f(a_1, \dots, a_n)$ to express |mapMerge| efficiently.
This array combinator is part of Z3's extended array theory \cite{DBLP:conf/fmcad/MouraB09} and axiomatized as
$\forall i.\,\text{map}_f(a_1, \dots, a_n)[i] = f(a_1[i], \dots, a_n[i])$.
While the combinator can in practice only be used with built-in functions, this is sufficient for our purposes:
Given Stainless' encoding of sets and maps, one can use the if-then-else function $\text{ite}$ of Z3, and translate |s.mapMerge(m1, m2)| as
$\text{map}_\text{ite}([\text{\lstinline{s}}], [\text{\lstinline{m1}}], [\text{\lstinline{m2}}])$.

\section{Experience}
\newcommand{\sparagraph}[1]{\smallskip\noindent \textbf{#1}\ }

We used our system to verify a number of benchmarks ranging in size and complexity.
Among the examples we developed are both shallowly and deeply mutable data structures, a model of an object allocator, and a parallelization primitive for the fork-join model.
Below we discuss our experience using the tool.

\sparagraph{Shallowly-Mutable Data Structures.}
We first consider ``shallowly-mutable'' data structures such as |Cell[T]| seen in Section~\ref{sec:example} whose mutable data is stored directly in its fields,
i.e., without any indirection.
They provide a simple baseline for our system and play an important role as building blocks for larger data structures such as trees and arrays with fine-grained separation properties.
However, shallowly-mutable data structures are useful in their own right:
For instance, we implemented |UpCounter| which tracks a monotonically increasing variable and maintains an invariant relative to the counter's initial value.
We also implemented a simple array (|SimpleArray|) and stack (|SimpleStack|) which essentially act as wrappers around functional data structures in that they only store
the reference to the head of an immutable list.
For instance, |SimpleArray[T]| consists of a single mutable field |var data: List[T]|.
In our examples we show safety wrt.\ bounds checks and non-emptyness when popping an element off the stack.
We found that our system easily deals with this kind of mutability, in particular since the associated operations typically require no recursion through stateful functions,
making them straightforward to verify and invalidate with counter-examples.

\sparagraph{Mutable Linked Lists and Queues.}
As an example of a more complex data structure we implemented multiple variations of a mutable, acyclic singly-linked list.
We focussed on an |append| operation, which takes two valid linked lists |l1| and |l2| with disjoint representations and concatenates them, leaving |l1| in a valid state.
This is challenging, since establishing the well-formedness of lists (e.g., the absence of cycles) requires knowledge of heap separation and an inductive proof that
maintains the property for intermediate nodes.

We considered several options to track a node's representation |repr|.
One could express |repr| as a recursive function as in Section~\ref{sec:example}, or, instead, as a mutable |@ghost| field on each node.
In our benchmarks we present two variants of the latter approach:
|MutList| encodes the ghost field |repr| as |List[AnyHeapRef]|, which has the added benefit of allowing predicates like |valid| to recurse on the representation, and
can be converted to a |Set[AnyHeapRef]| as required by our |reads| and |modifies| clauses.
|MutListSetsOnly| instead implements |repr| as |Set[AnyHeapRef]|, whose encoded form requires no further conversion to interact with the |mapMerge| primitive we use for framing.

We used a similar approach to implement |Queue|, which provides constant-time enqueue and dequeue methods using references to the first and last nodes.
Given a |valid| queue we prove that enqueue and dequeue are functionally correct with respect to a serialized representation similar to |toList| in Section~\ref{sec:example}.
The example demonstrates how safety properties can be established even in the presence of sharing and arbitrarily deep data structures.

\sparagraph{Slices and Monolithic Arrays.}
Arrays are one of the most common data structures found in imperative code and thus a worthwhile target for verification.
When specifying algorithms involving arrays it often pays to introduce slices, i.e., subarrays, as a means of abstraction.
By extending the |SimpleArray| example we arrived at |ArraySlice| which provides safe indexing, update and re-slicing operations wrt.\ an underlying array.
In the absence of sharing, this solution of encapsulating all array state in a single ``monolithic'' mutable heap object (the underlying array) is the natural and practical choice.
To analyze divide-and-conquer algorithms on arrays, on the other hand, we require some more fine-grained control, since we would like our dynamic frames to reflect the fact that
slices of an array may only access a subset of heap locations.

\sparagraph{Cell-Based Data Structures.}
A more complex representation based on lists of |Cell[T]|s allows us to achieve fine-grained framing of arrays and slices.
In example |CellArraySimple| we illustrate this approach and verify safety of accesses.
A more elaborate example, |CellDataStructuresAndRepr|, provides generalized infrastructure around splittable data structures.
We provide an interface |Repr| that, at its heart, requires implementers to provide a list of objects which make up the mutable representation of the data structure, and a proof of
uniqueness within that list.
Using |Repr| we can streamline proofs of heap separation for divide-and-conquer-style algorithms.
In this particular example, we implemented and verified the functional correctness of a |copy| operation on cell-based array slices.

\begin{figure}[btp]
\begin{lstlisting}[numbers=left,basicstyle={\footnotesize\sffamily\ourlstcolor}]
abstract class Task {
  @ghost def readSet: Set[AnyHeapRef]
  @ghost def writeSet: Set[AnyHeapRef] = { ??? } ensuring (_ $\subseteq$ readSet)

  def run(): Unit = { reads(readSet); modifies(writeSet); ??? : Unit }
}

def parallel(task1: Task, task2: Task): Unit = {
  reads(task1.readSet ++ task2.readSet)
  modifies(task1.writeSet ++ task2.writeSet)
  require((task1.writeSet $\cap$ task2.readSet == $\emptyset$) &&
          $\hspace{-0.13em}$(task2.writeSet $\cap$ task1.readSet == $\emptyset$))
  task1.run(); task2.run() // task1 and task2 complete before this function returns
}
\end{lstlisting}
  \caption{An interface for asynchronous computations and a sequential specification for fork-join parallelism.
  The {\protect \lstinline{???}} denotes unimplemented code in abstract classes.}
  \label{code:taskparallel}
\end{figure}

\sparagraph{Fork-Join Parallelism.}
Since dynamic frames in our system are simply given by read-only expressions, users may define their own imperative abstractions.
For instance, in |TaskParallel| we demonstrate how one can specify a primitive modelling fork-join parallelism.
Figure~\ref{code:taskparallel} shows an excerpt introducing the |Task| interface that encapsulates an asynchronous computation and declares the set of heap objects
that may be read and modified in the process.
Further below we define the |parallel(t1, t2)| construct itself, imposing a number of restrictions: Firstly, callers of |parallel| have to establish accessibility to
both |t1| and |t2|'s frames (lines 9-10). Secondly, we require that the read set of |t1| is disjoint from |t2|'s write set and vice-versa (lines 11-12).
This separation property justifies replacing our sequential model of |parallel| by a more efficient runtime implementation executing the two tasks concurrently.
Users can define new asynchronous tasks by implementing |Task|.
Operations such as those on cell-based data structures discussed above are straightforward to parallelize in this way.
Our introductory example (Section~\ref{sec:example}) could
be parallelized by defining a new class |TMapTask[T](t: Tree[T], f: T $\RA$ T)|
whose |run| method calls |tmap|, and replacing the recursive calls in |tmap| by |parallel(TMapTask(left, f), TMapTask(right, f))|.

Based on this experience, we are confident that our tool brings substantial value to the practice of verification of Scala programs.

%
%
{\raggedright
\bibliographystyle{splncs04}
\bibliography{ms}
}
\end{document}